\documentclass[pre,%
 reprint,
 amsmath,amssymb,
 aps,
]{revtex4-2}

\usepackage{graphicx}
\usepackage{dcolumn}
\usepackage{bm}
\usepackage{xcolor}

\begin{document}

\preprint{APS/123-QED}

\title{Memory-induced active particle ratchets: Mean currents and large deviations}

\author{Venkata D. Pamulaparthy}
\email{venkata.pamulaparthy.22@ucl.ac.uk}
\author{Rosemary J. Harris}%
 \email{rosemary.j.harris@ucl.ac.uk}
\affiliation{%
 University College London, Department of Mathematics \\ Gower Street,
London WC1E 6BT
}%




\date{\today}

\begin{abstract}
We analyse a continuous-time random walk model with stochastic reversals of direction. There is no external potential but the reorientation mechanism generates a nonzero current from asymmetry in the forward and backward waiting-time distributions (even when they have the same mean); the system \textcolor{black}{can therefore be considered} as a type of active particle ratchet. We derive an explicit expression for the mean ratchet current with exponentially distributed reorientation times and also develop a general renewal-theory framework to obtain the full large deviations, using this to  comment on the possibility of dynamical phase transitions.  
\end{abstract}

\maketitle


\section{\label{sec:level1} Introduction}
Systems that appear superficially symmetric but nonetheless generate a non-zero mean current are known as ratchets
 \cite{Feynman_ratchet, brownianmotors, DENISOV201477}. Many such models involve a particle with symmetric dynamics moving in a fluctuating external potential such as the sawtooth of the  paradigmatic flashing and rocking ratchet \cite{magnasco, rocking_ratchet, flashing_ratchet_review}. Recently, similar studies have been carried out with ``active" dynamics, e.g., run-and-tumble particles \cite{Angelani_2011, ratchet_run_and_tumble, fluctuation_3}. In this vein, we here present another type of active ratchet which removes the need for external potentials. Instead, in our model, the memory pertaining to the history of the process itself creates an asymmetry in the fluctuations, which are rectified to generate a non-zero current. Specifically, we work with semi-Markov dynamics where memory is given by non-exponential waiting-time distributions, meaning that the probability of transitions from a given state depends on how long the system has waited at that state.



Our model was briefly introduced in \cite{Pamulaparthy_2025} but analytical calculations there were limited to non-Markovian cases for which equivalent Markov representations with hidden states are possible, the  main focus being the validation of the reinforcement learning method described in that work. In the present paper, we are concerned with deeper theoretical examination of the model, covering more general cases with light-tailed and heavy-tailed waiting-time distributions, including those for which straightforward hidden-Markov representations are not possible.

In this work, apart from giving an explicit expression for mean ratchet current, we further provide a method to calculate the associated fluctuations, noting that such fluctuations have already attracted interest in more conventional ratchet models \cite{fluctuation1,fluctuation_2}. We inspect current fluctuations through the lens of large-deviation theory \cite{Ellis2006,Derrida_2007,touchette1}, in which a key quantity is the so-called \textit{scaled cumulant generating function} (SCGF). The SCGF is of central importance in this paper; as we shall see it enables analysis of trends away from the mean and examination of the possibility of dynamical phase transitions.

The study of ratchet currents and fluctuations is not merely of theoretical interest. For example, ratchets of various types have been extensively used to model molecular motors \cite{ ratchet_molecular_motors, Angelani_2, fluctuation_2, ratchet_dna_molecular_motor, memory_induced_ratchet}. For our model, a particular motivation comes from  the relation to run-and-tumble type systems, which are popularly used for describing biological transport, e.g., bacterial motion \cite{chemotaxis2,chemotaxis1, run_and_tumble_nature}. Furthermore, memory could play a significant role in modelling realistic dynamics, as suggested in \cite{non_markov_run_and_tumble2}; see also \cite{Generalize_run_and_tumble} for a broad treatment of generalized run-and-tumble processes and applications.

We now provide a brief overview of this manuscript. In Sec.~\ref{Model}, we develop the model and the associated mathematical notation for analysis. In Sec.~\ref{typeA ratchet}, we describe examples of ratchet models with phase-type waiting times, where hidden-Markov representations are possible, allowing for verification of means and large deviations computed using the methods in this paper. In Sec.~\ref{mean_current_section}, finally equipped with all the necessary mathematical background, we describe how directional preference arises to form ratchet currents. We derive formulas for computing the mean ratchet current and discuss some properties. In Sec.~\ref{ratchet_large_deviations}, we present computations for the SCGF of ratchet currents via  a framework using renewal theory and Laplace transforms. 
We then use the SCGF to reveal the possibility of dynamical phase transitions in certain types of exotic ratchets. In Sec.~\ref{discussion}, we conclude with a brief discussion. 

\textcolor{black}{It turns out that our work is closely related to the recent significant contribution of  Santra et al.~\cite{Santra}, in which a similar approach is used to study the dynamics of switching processes with calculation of the spatial cumulants (in particular, the effective diffusion coefficient). However, our setup is somewhat different to the examples considered by those authors and we have a complementary focus on the ratchet effect and the full large deviation calculation.   
}

\section{\label{Model}Model}

We consider the archetypal framework of continuous-time random walks \textcolor{black}{(CTRWs)} \cite{CTRW}, modified to include random flips in particle orientation.  The dynamics may be described as a composite process in a clockwise channel and an anti-clockwise channel with a reorientation mechanism for switching between the two channels \footnote{\textcolor{black}{Such two-channel models, in fact, have a long history; for example, see \cite{Schlesinger_two_lane} where  the mean and diffusivity of a related temporally correlated random walk are calculated in the context of an application to superionic conductivity. At least for some special cases, there may also be a connection to the more recent concept of lifted Markov chains popularized in computer science \cite{lifted_markov_chains}.}}. The model can be thought of as a run-and-tumble process with forward and backward runs separated by tumbles. Here, we impose periodic boundary conditions to clearly demonstrate calculations in a finite-state space. However, since we are only interested in current statistics, the results should also apply for a particle moving in an infinite two-channel system. 

\begin{figure}
\centering
\includegraphics[scale=0.75]{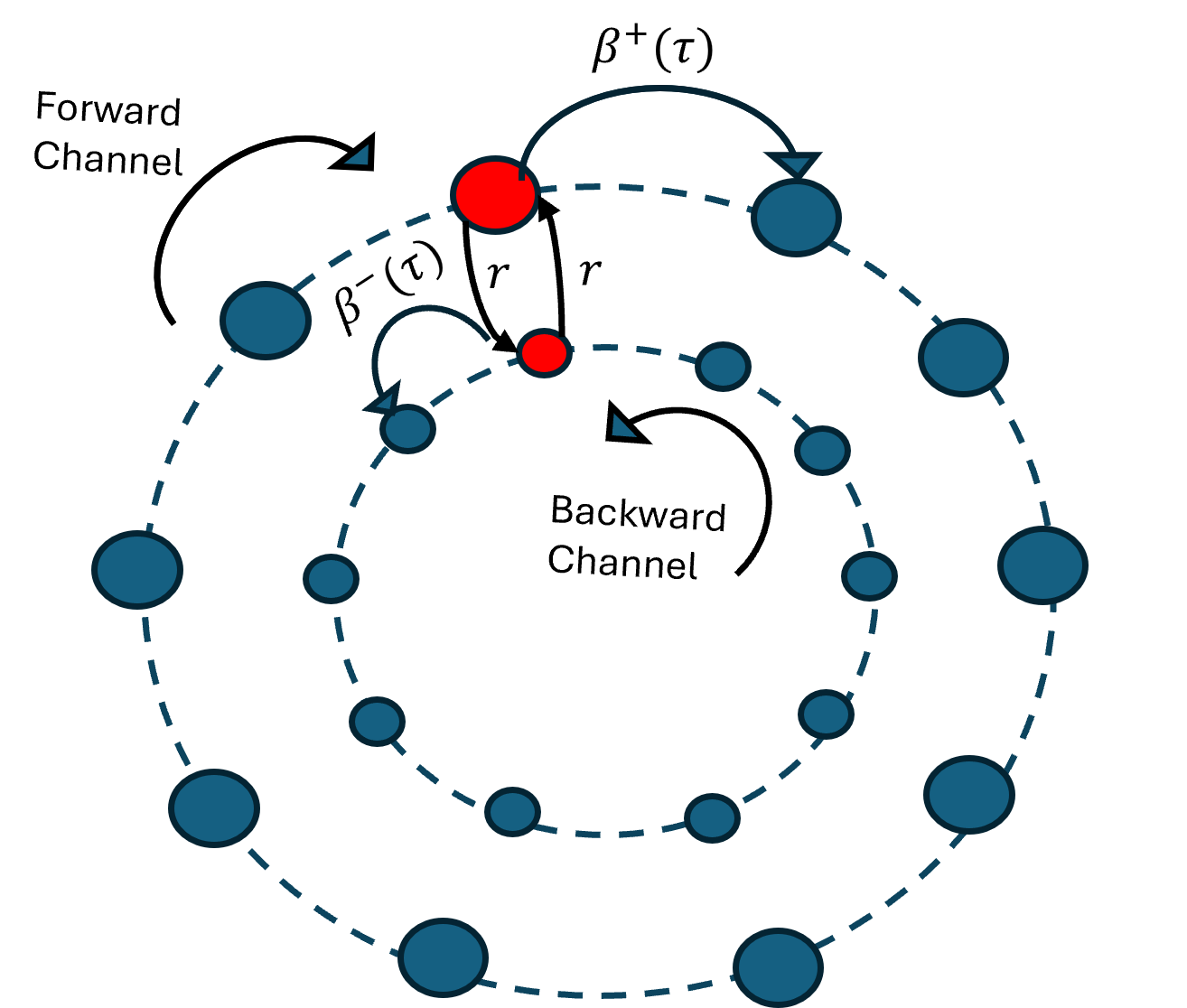}
\caption{\label{two_channel_fig}Two-channel \textcolor{black}{CTRW} on a ring. 
The particle transitions are restricted to the respective channels with rates $\beta^{+}(\tau)$ and $\beta^{-}(\tau)$ until a reorientation occurs with rate $r$ and the particle switches channels.}
\end{figure}

The system is visualized in Fig.~\ref{two_channel_fig}.
In the absence of reorientations, the time spent at a given site is sampled from a waiting-time distribution of $\psi^{+}(t)$ in the forward channel and $\psi^{-}(t)$ in the backward channel, after which the particle transitions to the next site in that channel. The waiting-time distributions obey the normalization condition $\int^{\infty}_{0}\psi^{+}(\tau ) \mathrm{d}\tau=1$ and $\int_{0}^{\infty}\psi^{-}(\tau)\mathrm{d}\tau=1$. If the two densities are non-exponential, the corresponding rates, denoted here by $\beta^{+}(\tau)$ and $\beta^{-}(\tau)$, depend in general on the waiting times $\tau$; we have
\begin{equation}\label{4_ratchet}
\beta^{+}(\tau) =\frac{\psi^{+}(\tau)}{\varphi^{+}(\tau)} \ , \quad \beta^{-}(\tau)= \frac{\psi^{-}(\tau)}{\varphi^{-}(\tau)},
\end{equation}
with survival functions $\varphi^{+}(\tau)=1-\int_{0}^{\tau}\psi^{+}(t)\mathrm{d}t$ and  $\varphi^{-}(\tau)=1-\int_{0}^{\tau}\psi^{-}(t)\mathrm{d}t$. 
The dependence of the rates on waiting times gives rise to memory in the system. In particular, on including the reorientation mechanism, the probabilities describing whether the particle jumps to the next site in the same channel or switches to the other channel are dependent on how long the particle has waited at the current site.
To simplify analysis, we assume unless stated otherwise, that the flips in particle orientation are prescribed by an exponential waiting-time distribution with a constant rate $r$. \textcolor{black}{The process is somewhat reminiscent of reset systems such as described in \cite{Evans_resetting}, but with the distinction that the reset here is in the clock variable (when the particle changes lane) and not the position. For spatial resets as in \cite{Evans_resetting}, models typically reach a well-defined stationary state due to the particle being returned to the initial position or distribution. In our case, although there is no spatial resetting, the system still generically has a stationary state as the particle dynamics is considered on a ring.} 

A natural question to ask here is if the system has a directional preference, determined by some non-zero current at  long times. We are particularly interested in scenarios where the distributions $\psi^{+}(\tau)$ and $\psi^{-}(\tau)$ are different, but have the same mean,  i.e., $\int_{0}^{\infty} \tau \psi^{+}(\tau) \mathrm{d}\tau = \int_{0}^{\infty} \tau \psi^{-}(\tau) \mathrm{d}\tau $. If the current is non-zero despite the means being the same, then we can call the system a ratchet. The study of the mean and fluctuations of such ratchet currents is the main subject of this paper. However, we first develop some required mathematical machinery.

We formalize the aforementioned dynamics by tracking the particle transitions on a ring with $L$ sites, i.e., on a state space $\mathcal{S}$ of $L$ positional vectors: $\{|1\rangle,|2\rangle,|3\rangle,\cdots,|L\rangle\}$.  To be concrete, if the particle is at the $n^\mathrm{th}$ site on the ring, it is represented by the vector $|n\rangle$ whose $n^{\mathrm{th}}$ element is one while all other elements are zero.

To include the chirality, i.e., the channel structure, we employ representations used in quantum walks \cite{quantum_rw}. However, we emphasize that our system is purely classical and the quantum formulation is used only for notational convenience. Specifically, we define vectors
\begin{equation}
| \uparrow \rangle = \begin{bmatrix} 1 \\ 0 \end{bmatrix}, \quad | \downarrow \rangle = \begin{bmatrix} 0 \\ 1 \end{bmatrix},
\end{equation}
to indicate if the particle is in the forward or the backward channel respectively. In this new representation, we can write the particle states in the two-channel process as
\begin{equation}\label{1_ratchet}
|\mathbf{n}^{+}\rangle  =  |\uparrow\rangle \otimes |n\rangle, \quad  
|\mathbf{n}^{-}\rangle = |\downarrow \rangle \otimes |n\rangle.
\end{equation}
 Here, $|\mathbf{n}^{+}\rangle$ indicates the particle is at the position $|n\rangle$ on the ring in the forward channel while $|\mathbf{n}^{-}\rangle$ indicates that the particle is at the same position but in the backward channel. These vector objects belong to the extended state space $\mathcal{S}^{\mathrm{e}}$ of $2L$ elements: $\{|\mathbf{1}^{+}\rangle,|\mathbf{2}^{+}\rangle,|\mathbf{3}^{+}\rangle,\dots,|\mathbf{L}^{+}\rangle,|\mathbf{1}^{-}\rangle,|\mathbf{2}^{-}\rangle,|\mathbf{3}^{-}\rangle,\dots,|\mathbf{L}^{-}\rangle \}$. Note that the elements of $\mathcal{S}^{e}$ are distinguished from the elements of $\mathcal{S}$ by using bold notation. Our ultimate aim is to obtain the mean current in the position space $\mathcal{S}$ but, to this end, it is convenient to work with the extended state space $\mathcal{S}^{\mathrm{e}}$. We emphasize that, even in this extended state space the process is non-Markovian; specifically, a semi-Markov process with memory introduced via the non-exponential nature of the densities $\psi^{+}(\tau)$ and $\psi^{-}(\tau)$.

In the notational framework now established, we write the conditional waiting-time densities for transitions in the two channels as
\begin{align}
& W_{|\mathbf{n^{+}}\rangle \to |(\mathbf{n+1})^{+}\rangle}(\tau) = \psi^{+}(\tau) e^{-r\tau} 
\end{align}
and
\begin{align}
& W_{|\mathbf{n^{-}}\rangle \to |\mathbf{(n-1)}^{-}\rangle}(\tau) = \psi^{-}(\tau)e^{-r\tau}.
\end{align}
Similarly, we have the waiting-time densities for switching between the channels
\begin{align}
W_{|\mathbf{n^{+}}\rangle \to |\mathbf{n}^{-}\rangle }(\tau) = re^{-r\tau} \varphi^{+}(\tau)
\end{align}
and
\begin{align}
W_{|\mathbf{n^{-}}\rangle \to |\mathbf{n}^{+}\rangle}(\tau) = r e^{-r\tau} \varphi^{-}(\tau).
\end{align}
 The unconditional waiting-time distributions for any move out of a given site, either a transition within the channel or a reorientation, are then given by
\begin{align}\label{2_ratchet}
  W_{|\mathbf{n^{+}}\rangle}(\tau) = (\psi^{+}(\tau) + r \varphi^{+}(\tau)) e^{-r \tau}
 \end{align}
 and
 \begin{align}
  W_{|\mathbf{n}^{-}\rangle}(\tau) = (\psi^{-}(\tau) + r \varphi^{-}(\tau)) e^{-r \tau}.
\end{align}
Intuitively, in each case the first term is the conditional distribution given that a channel transition succeeds and the reorientation fails in time $\tau$, while the second term is the conditional distribution given that a reorientation occurs instead and the particle switches to the opposite channel.

It is also helpful to consider the jump probabilities giving the mass distribution over target sites given a jump out of a particular site occurs after a waiting-time $\tau$. These probabilities are given in terms of the rates; when the particle is in the forward channel, we have
 \begin{align}\label{forward_transitions}
  p_{|\mathbf{n^{+}}\rangle \to |(\mathbf{n+1})^{+}\rangle} (\tau) = \frac{\beta^{+}(\tau)}{\beta^{+}(\tau )+ r} 
\end{align}
and
\begin{align}
p_{  |\mathbf{n}^{+}\rangle  \to |\mathbf{n}^{-}\rangle }(\tau) = \frac{r}{\beta^{+}(\tau )+ r},
\end{align}
while, when the particle is in the backward channel, we have
\begin{align}
\label{backward_transitions}
 p_{ |\mathbf{n}^{-}\rangle   \to |(\mathbf{n-1})^{-} \rangle} (\tau) = \frac{\beta^{-}(\tau)}{\beta^{-}(\tau) + r}
\end{align}
and
\begin{align}
p_{|\mathbf{n}^{-}\rangle  \to |\mathbf{n}^{+}\rangle}(\tau) = \frac{r}{\beta^{-}(\tau )+ r}.
\end{align}

\section{Phase-type examples}
\label{typeA ratchet}
Having introduced the general structure of the two-channel model, we here present specific examples with waiting times prescribed by phase-type distributions \cite{Cox_1955_phase_type,phase-type-theory}. We emphasize that the interest of this paper is not restricted to such models but they are useful test cases since semi-Markov systems with phase-type waiting-time distributions have equivalent Markov representations allowing for calculation of the SCGF via spectral analysis of the tilted generator, as already discussed in \cite{Pamulaparthy_2025}. The examples of this section will serve as a reference throughout this paper for verifying the mean and fluctuations of ratchet currents, as predicted by our general methods. To simplify calculations and elucidate the underlying mechanisms, we here consider examples of two-channel models where waiting-time distributions are non-exponential only in one channel.

\subsection{Hypoexponential-exponential ratchet}
 
As a first example, we set the forward waiting-time distribution, ${\psi}^{+}(\tau)$, as a two-parameter hypoexponential distribution,
\begin{equation}
\psi^{+}(\tau)=\frac{p_1p_2}{p_1-p_2} (e^{-p_2\tau}-e^{-p_1\tau}),
\end{equation}
where we have in general $\ p_1 \neq p_2$ and the mean of the distribution is \textcolor{black}{$(p_{1}+p_{2})/p_{1}p_{2}$}. This is an example of a phase-type distribution obtained by convolving exponential distributions with different rates. [The special case $p_1=p_2=p$ gives the gamma distribution, $p^{a} t^{a-1}e^{-p \tau}/{\Gamma{(a)}}$, with rate parameter $p$ and shape parameter $a=2$.] The two-state hypoexponential subprocess then has a \textit{fine-grained} hidden-Markov representation consisting of two states in series having exponentially distributed waiting times with rates $p_1$ and $p_2$. We assume that the backward subprocess has an exponentially distributed waiting-time distribution with rate $q$, given by $\psi^{-}(\tau)=q e^{-q\tau}$. For the waiting-time distributions in the forward and backward subprocesses to have equal mean, we require that the parameters follow the relation
\begin{equation}
\label{equal_means_hypo}
\frac{p_1p_2}{p_1+p_2}  = {q}.
\end{equation}
We call this the hypoexponential-exponential ratchet.

\begin{figure}
\includegraphics[scale=0.75]{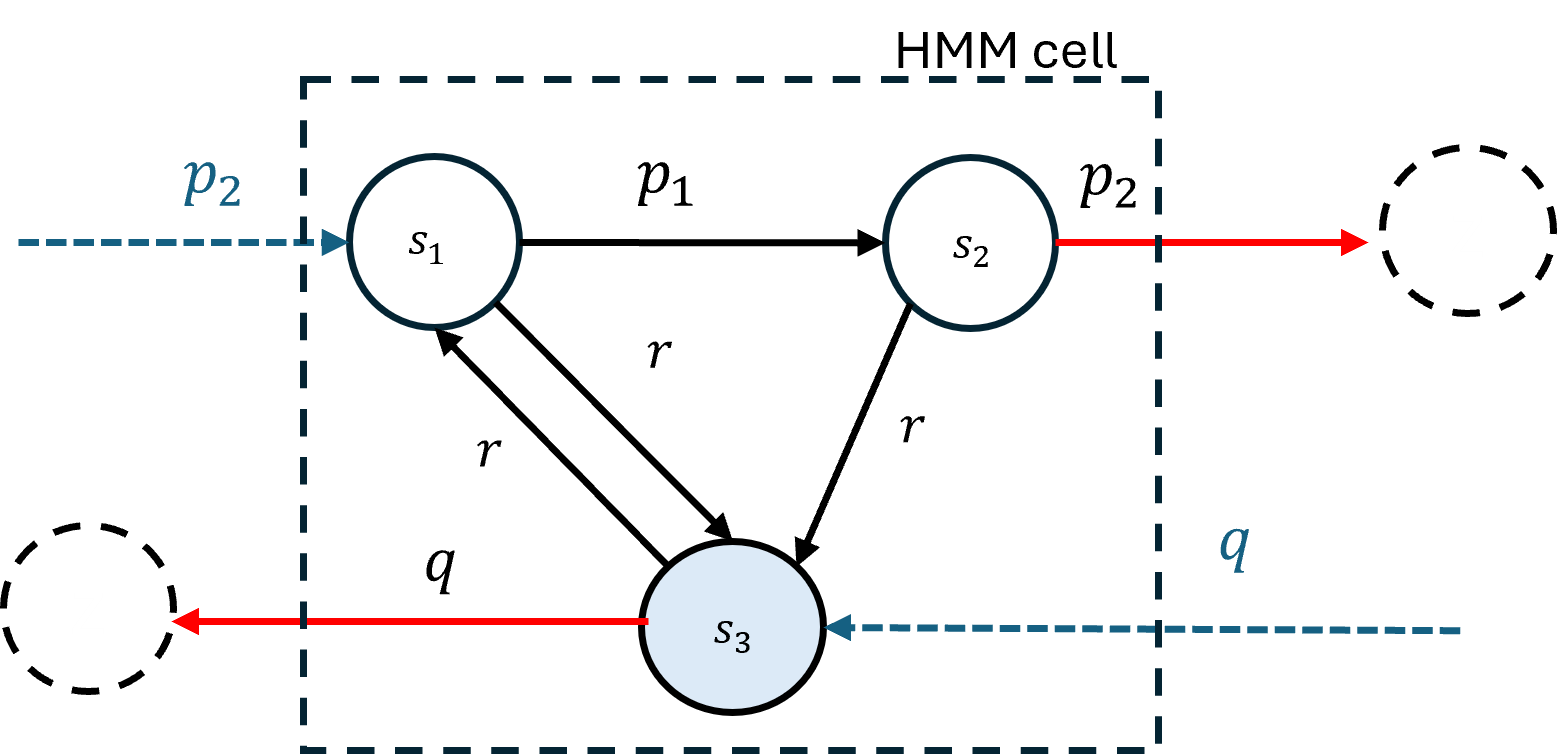}
\caption{HMM cell structure for a hypoexponential-exponential ratchet. The dashed box shows the hidden sites at one position on the ring.}
\label{ratchet_hmm_hypo_figure}
\end{figure}

The equivalent hidden-Markov model (HMM) for this system is given in Fig.~\ref{ratchet_hmm_hypo_figure}. 
Each site in the periodic hypoexponential-exponential ratchet corresponds to a cell with three hidden states $s_1$, $s_2$ and $s_3$. The forward run of the ratchet is represented by Markov transitions between states $s_1$ and $s_2$, connected in \emph{series} with rates $p_1$ and $p_2$, yielding an overall hypoexponential waiting-time distribution. The backward run is Markovian (no additional hidden states needed) with a transition out of $s_3$ to the next state in the backward channel occurring with a rate $q$. Inter-channel moves (reorientations) all occur with rate $r$; the particle can transition from $s_3$ to $s_1$ and return from either $s_1$ or $s_2$ to $s_3$. 

\subsection{Hyperexponential-exponential ratchet}

Next, we take the forward waiting-time distribution as a two-parameter hyperexponential distribution, with equiprobable branches and rates in each branch being $p_1$ and $p_2$ respectively, given by the probability density function
\begin{equation}
\psi^{+}(\tau)=\frac{1}{2}(p_{1}e^{-p_{1}\tau}+ p_{2} e^{-p_{2}\tau}).
\end{equation}
Again, the waiting-time distribution of the backward subprocess, $\psi^{-}(\tau)$ is assumed to be exponential with rate $q$.  We then have the condition,
\begin{equation}
\label{equal_mean_hyper}
\frac{2p_1 p_2}{p_1+p_2} = q 
\end{equation}
for equal means. We call this the hyperexponential-exponential ratchet.  

\begin{figure}
\includegraphics[scale=0.75]{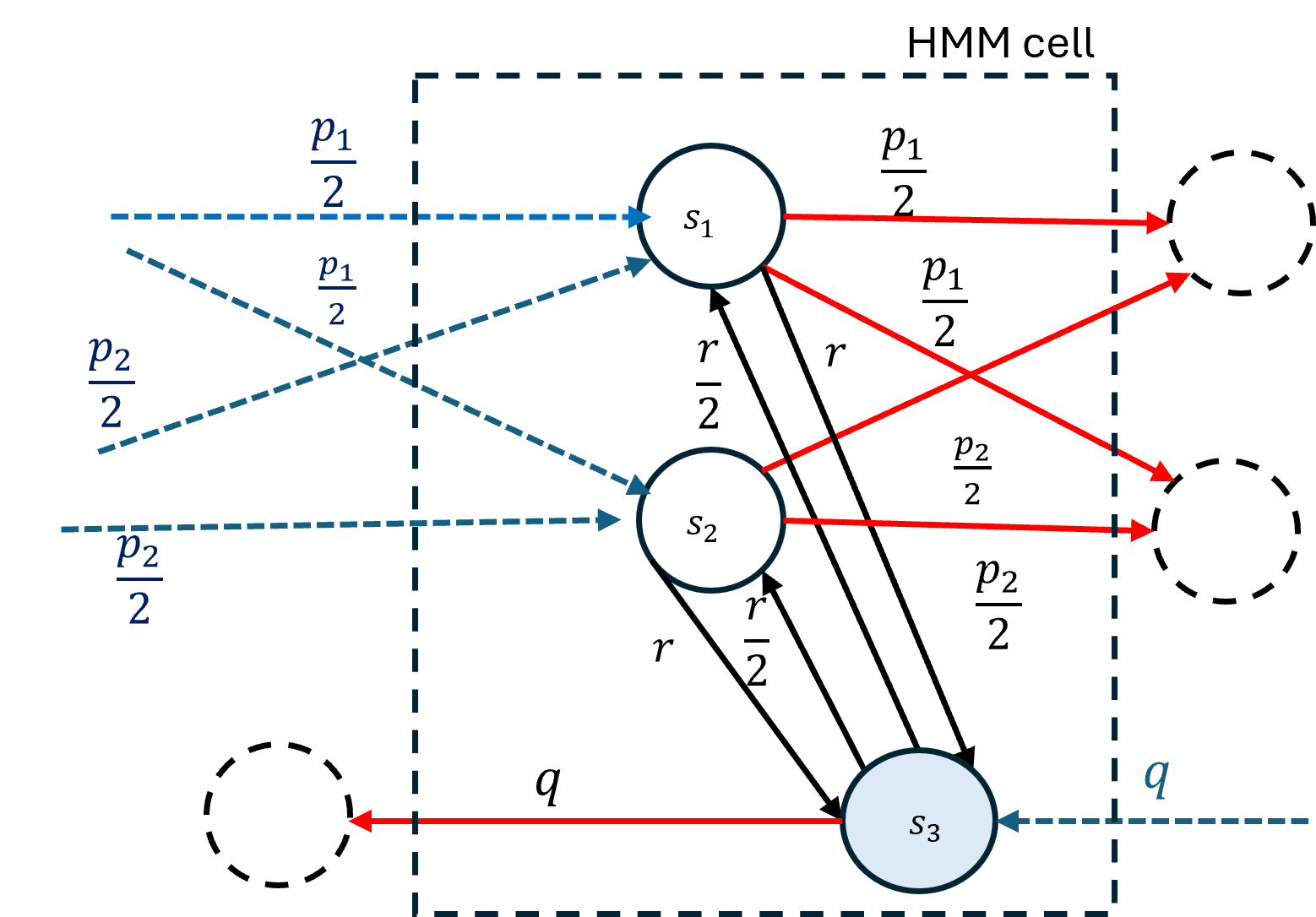}
\caption{HMM cell structure for a hyperexponential-exponential ratchet. The dashed box shows the hidden sites at one position on the ring.}
\label{ratchet_hmm_hyper_figure}
\end{figure}

The equivalent HMM for the hyperexponential-exponential ratchet is shown in Fig.~\ref{ratchet_hmm_hyper_figure}. 
The corresponding cell structure, as in the hypoexponential-exponential case, has three hidden sites $s_1$, $s_2$ and $s_3$. However, here the forward run contains hidden states $s_1$ and $s_2$ connected in \emph{parallel} over equiprobable branches, leading to a hyperexponential distribution. The backward run is again described by Markovian transitions out of $s_3$ with rate $q$. In this case, when the particle is in the backward run, inter-channel moves (reorientations) from $s_3$ to \emph{both} $s_1$ and $s_2$ are possible each with rate $r/2$. During the forward run the particle returns to $s_3$ from either $s_1$ or $s_2$ with rate $r$.

\section{Mean current}
\label{mean_current_section}

We now turn our focus to the derivation and analysis of the ratchet current. In the context of the ratchet system, the time-integrated current $J(t)$ is defined as the difference, $J^{+}(t)-J^{-}(t)$, between the total number of forward and backward jumps in the interval $(0,t]$, accumulated during the time spent in forward and backward channels respectively. We are interested in the statistics of the time-averaged current $j(t) = J(t)/t$, focusing on the calculation of the mean current $\langle j\rangle$ in this section and considering the fluctuations later in Sec.~\ref{ratchet_large_deviations}. Note that we use angular brackets  throughout this paper to denote averaging of quantities over the stationary state.

\subsection{Generalized master equation}

One way to derive the mean current is via the generalized master equation (GME) \cite{espacito_ctrw, GeneralizedME} for semi-Markov jump processes on the state space $\mathcal{S}^{\mathrm{e}}$. Closely following \cite{espacito_ctrw}, we start by writing the GME for the time-evolution of the probability $\rho_{|\mathbf{m}\rangle}(t)$ of being in any state $|\mathbf{m}\rangle \in \mathcal{S}^{\mathrm{e}}$:   
\begin{align}
\label{GME_ratchet}
 \nonumber \dot{\rho}_{|\mathbf{m}\rangle}(t)= I(t) + \sum_{|\mathbf{m'}\rangle} \int_{0}^{t} & \mathrm{d}\tau(\kappa_{|\mathbf{m'}\rangle \to |\mathbf{m}\rangle}(\tau)\rho_{|\mathbf{m'}\rangle}(t-\tau)  \\ - &\kappa_{|\mathbf{m}\rangle \to |\mathbf{m'}\rangle}(\tau) \rho_{|\mathbf{m}\rangle}(t-\tau)),
\end{align}
 where $\kappa_{|\mathbf{m'}\rangle \to |\mathbf{m}\rangle}(\tau)$ and $\kappa_{|\mathbf{m}\rangle \to |\mathbf{m'}\rangle}(\tau)$ represent memory kernels and $I(t)$ pertains to the initial conditions. 
 
It is useful to consider the GME in the Laplace-transformed space, as shown in \cite{espacito_ctrw}. We find that the Laplace-transformed memory kernel is given by
\begin{align}
\tilde{\kappa}_{|\mathbf{m'}\rangle \to |\mathbf{m}\rangle}(\nu)= \frac{\tilde{W}_{|\mathbf{m'}\rangle \to |\mathbf{m}\rangle}(\nu)}{\tilde{\varphi}_{|\mathbf{m'}\rangle}(\nu)},
\end{align}
where $\tilde{W}_{|\mathbf{m'}\rangle \to |\mathbf{m}\rangle}(\nu)$ is the Laplace transform of the conditional waiting-time distribution $W_{|\mathbf{m'}\rangle \to |\mathbf{m}\rangle}(\tau)$ and $\tilde{\varphi}_{|\mathbf{m'}\rangle}$ is the Laplace transformed survival function for staying in $|\mathbf{m'}\rangle$.
In the notation from Sec.~\ref{Model}, for transitions in the forward channel we have
\begin{align}\label{kap_forward}
\nonumber \tilde{\kappa}_{|\mathbf{m^{+}}\rangle \to  |(\mathbf{m+1})^{+}\rangle}(\nu) &=  \frac{\int_{0}^{\infty} e^{-\nu \tau}\psi^{+}(\tau)e^{-r\tau} \mathrm{d}\tau}{\int_{0}^{\infty} e^{-\nu \tau}\varphi^{+}(\tau)e^{-r\tau} \mathrm{d}\tau} \\ &=\frac{\tilde{\psi}^{+}(\nu+r)}{\tilde{\varphi}^{+}(\nu+r)}
\end{align}
and for transitions in the backward channel we have
\begin{align}
\label{kap_backward}
\nonumber \tilde{\kappa}_{|(\mathbf{m}^{-}\rangle \to  |(\mathbf{m-1})^{-}\rangle}(\nu)&= \frac{\int_{0}^{\infty} e^{-\nu \tau}\psi^{-}(\tau)e^{-r\tau} \mathrm{d}\tau}{\int_{0}^{\infty} e^{-\nu \tau}\varphi^{-}(\tau)e^{-r\tau} \mathrm{d}\tau}\\&=\frac{\tilde{\psi}^{-}(\nu+r)}{\tilde{\varphi}^{-}(\nu+r)}.
\end{align}
 The above equations use the shifting property of Laplace transforms: $\int_{0}^{\infty} e^{-\nu \tau} e^{-r\tau} f(\tau) \mathrm{d}\tau = \tilde{f}(\nu+r)$. For transitions between channels, the kernels reduce to a Markovian rate 
\begin{align}
 \tilde{\kappa}_{|\mathbf{m^{+}}\rangle \to |\mathbf{m}^{-}\rangle}(\nu)= \tilde{\kappa}_{|\mathbf{m^{-}}\rangle \to |\mathbf{m}^{+}\rangle}(\nu)= r.
\end{align}

The term $I(t)$ can also be defined in the Laplace domain:
\begin{align}
\tilde{I}(\nu) =  \nonumber \sum_{|\mathbf{m'}\rangle} & \{ \tilde{\kappa}_{|\mathbf{m'}\rangle \to |\mathbf{m}\rangle}(\nu)  [\tilde{\varphi}_{|\mathbf{m'}\rangle}(\nu) -\tilde{\Phi}_{|\mathbf{m'}\rangle}(\nu)] \rho_{|\mathbf{m'}\rangle}(0)\\ & -\tilde{\kappa}_{|\mathbf{m}\rangle \to |\mathbf{m'}\rangle}(\nu) [\tilde{\varphi}_{|\mathbf{m}\rangle}(\nu)- \tilde{\Phi}_{|\mathbf{m}\rangle}(\nu)] \rho_{|\mathbf{m}\rangle}(0) \}.
\end{align}
Here $\tilde{\Phi}_{|\mathbf{m'}\rangle}(\nu)$ is the Laplace transform of the survival function ${\Phi}_{|\mathbf{m'}\rangle}(t) =1-\int_{0}^{t}\Psi_{|\mathbf{m'}\rangle}(\tau)\mathrm{d}\tau$, where $\Psi_{|\mathbf{m'}\rangle}(t)$ denotes the special waiting-time density describing the time to the first departure from $|\mathbf{m'}\rangle$ after the measurement of the process began at $t=0$. In the Markov case, by definition, there is no history dependence, implying that $\Phi_{|\mathbf{m}\rangle}(t)=\varphi_{|\mathbf{m}\rangle}(t)$ and we always have $I(t)=0$. However, for non-Markov systems, including the semi-Markov dynamics depicted here, $\Psi_{|\mathbf{m}\rangle}(t)$  generally depends on the history of the process before $t=0$ and $I(t)$ can be non-zero.

To determine the long-time properties of the mean currents we need only consider the asymptotics of the semi-Markov GME. Ergodic semi-Markov processes, guaranteed when the conditional waiting-time densities $W_{|\mathbf{m'}\rangle \to |\mathbf{m}\rangle}(\tau)$  have finite moments, are absorbed at long times into well-defined stationary dynamics \cite{Asympt_CTRW}. In this case, the term  $I(t)$ is suppressed (i.e., we have  $\lim_{t \to \infty}I(t) = 0$) \cite{espacito_ctrw} and the memory kernels are given by
\begin{equation}
\tilde{\kappa}_{\mathbf{|m'\rangle} \to \mathbf{|m\rangle}}(0) = \frac{\tilde{W}_{\mathbf{|m'\rangle} \to \mathbf{|m\rangle}} (0)}{\tilde{\varphi}_{\mathbf{|m'\rangle}}(0)}. 
\end{equation}
The GME of \eqref{GME_ratchet} thus simplifies in the asymptotic limit to an effective Markov master equation
\begin{equation}
\dot{\rho}_{\mathbf{|m\rangle}}(t) = \sum_{\mathbf{|m'\rangle}} \tilde{\kappa}_{\mathbf{|m'\rangle} \to \mathbf{|m\rangle}}(0) \rho_{\mathbf{|m'\rangle}}(t) - \tilde{\kappa}_{\mathbf{|m\rangle} \to \mathbf{|m'\rangle}}(0)\rho_{\mathbf{|m\rangle}}(t).
\end{equation}
To be concrete, the process may then be described by a $2L \times 2L$ effective Markov generator
\begin{align}
\label{effective_generator}
\nonumber
\tilde{\boldsymbol{\kappa}}(0) = & \begin{bmatrix} 1 & 0 \\ 0 & 0 \end{bmatrix} \otimes \boldsymbol{\tilde{\kappa}^{+}}(0)+ \begin{bmatrix} 0 & 0\\  0 & 1 \end{bmatrix} \otimes \boldsymbol{\tilde{\kappa}^{-}}(0) \\ 
& + \begin{bmatrix} -r & r \\ r & -r \end{bmatrix} \otimes   \mathrm{I}_{L \times L},
\end{align} 
where $\boldsymbol{\tilde{\kappa}^{+}}(0)$ and $\boldsymbol{\tilde{\kappa}}^{-}(0)$ are effective Markov generators of subprocesses describing  forward and backward unidirectional walks, 
with rates $\tilde{\kappa}_{ |\mathbf{m}^{+} \rangle \to  |\mathbf{(m+1)}^{+}\rangle}(0)$
and $\tilde{\kappa}_{|\mathbf{m}^{-} \rangle \to |(\mathbf{m-1})^{-}\rangle}(0)$
respectively, 
while $\mathrm{I}_{L \times L}$ represents an $L \times L$ identity matrix.

With the effective generator defined, we now turn to the familiar Markov stationary condition
\begin{equation}
\label{stationary_condition}
\sum_{\mathbf{|m\rangle}} \tilde{\kappa}_{\mathbf{|m'\rangle} \to \mathbf{|m\rangle}}(0) \rho^{\mathrm{ss}}_{\mathbf{|m'\rangle}} - \tilde{\kappa}_{\mathbf{|m\rangle} \to \mathbf{|m'\rangle}}(0)\rho^{\mathrm{ss}}_{\mathbf{|m\rangle}} = 0,
\end{equation}
which can be written more compactly as
\begin{equation}
\tilde{\boldsymbol{\kappa}}(0) | \boldsymbol{\rho}^{\mathrm{ss}}\rangle =  0.
\end{equation}
Here ${\rho}^{\mathrm{ss}}_{\mathbf{|m'\rangle}}$ and $\rho^{\mathrm{ss}}_{\mathbf{|m\rangle}}$ are the elements of the normalized stationary probability vector $|\boldsymbol{\rho}^{\mathrm{ss}}\rangle$, which is the principal eigenvector of $\boldsymbol{\tilde{\kappa}}(0)$. 
To construct  $|\boldsymbol{\rho}^{\mathrm{ss}}\rangle$ for the two-channel model,  we focus momentarily on the unidirectional subprocesses described by the generators $\tilde{\boldsymbol{\kappa}}^{+}(0)$ and $\boldsymbol{\tilde{\kappa}}^{-}(0)$. We can then introduce stationary probability vectors $|\boldsymbol{\rho^{+}}\rangle$ and $|\boldsymbol{\rho^{-}}\rangle$ for the forward and backward subprocesses which satisfy  
\begin{align}
\tilde{\boldsymbol{\kappa}}^{+}(0)|\boldsymbol{{\rho}^{+}}\rangle=0 
\end{align}
and
\begin{align}
\tilde{\boldsymbol{\kappa}}^{-}(0)|\boldsymbol{\rho^{-}}\rangle=0.
\end{align}
Now, we note that the symmetric matrix in the last term in \eqref{effective_generator} is the part of the generator that controls the reorientation. Physically, the symmetry means that the average time spent in the two channels is the same; there is an equal chance of finding the process in the forward or backward channel. Hence, we can decompose the stationary probability vector of the two-channel process as
\begin{equation}
\label{decomposition}
|\boldsymbol{\rho}^{\mathrm{ss}}\rangle = \frac{1}{2} (| \uparrow \rangle \otimes |\boldsymbol{\rho^{+}}\rangle +  | \downarrow \rangle \otimes  |\boldsymbol{\rho^{-}}\rangle).
\end{equation}

 We can now derive explicitly the mean current in this stationary state. Since each subprocess is ergodic and effectively Markovian in the long-time limit, the mean count of jumps in each channel is given by averaging the transition rates over the stationary probability vector. Thus, recalling that the effective forward and backward subprocesses are homogeneous unidirectional walks with rates   $\tilde{\kappa}_{|\mathbf{m^{+}}\rangle \to  |(\mathbf{m+1})^{+}\rangle }(0) = \tilde{\psi}^{+}(r)/\tilde{\varphi}^{+}(r)$  and  $\tilde{\kappa}_{|\mathbf{m}^{-}\rangle \to |(\mathbf{m-1})^{-}\rangle }(0)=\tilde{\psi}^{-}(r)/\tilde{\varphi}^{-}(r)$ respectively [see \eqref{kap_forward} and \eqref{kap_backward}], we obtain the average forward and backward currents in the system as
\begin{align}
\label{mean_j_{+}}
 \langle j^{+} \rangle  =  \frac{\tilde{\psi}^{+}(r)}{\tilde{\varphi}^{+}(r)} \langle\mathbf{1}|\boldsymbol{\rho^{+}}\rangle =\frac{\tilde{\psi}^{+}(r)}{\tilde{\varphi}^{+}(r)}
\end{align}
and
\begin{align}
\label{mean_j_{-}}
 \langle j^{-} \rangle & = \frac{\tilde{\psi}^{-}(r)}{\tilde{\varphi}^{-}(r)}   \langle\mathbf{1}|\boldsymbol{\rho^{-}}\rangle  =\frac{\tilde{\psi}^{-}(r)}{\tilde{\varphi}^{-}(r)}.
\end{align}
Here $\langle \mathbf{1} |$ is a row vector of all ones and we use the  normalization of probability vectors: $\langle \mathbf{1}|\boldsymbol{\rho}^{+}\rangle=1$ and $\langle \mathbf{1}|\boldsymbol{\rho}^{-}\rangle=1$. The mean current of the full system is then given by
\begin{align}
\label{mean_j+}
 \nonumber \langle j \rangle &=  \frac{1}{2} ( \langle j^{+} \rangle -\langle j^{-}\rangle)  \\ &= \frac{1}{2} \left( \frac{\tilde{\psi}^{+}(r)}{\tilde{\varphi}^{+}(r)} - \frac{\tilde{\psi}^{-}(r)}{\tilde{\varphi}^{-}(r)} \right),
\end{align}
where the coefficient $1/2$ emerges naturally from the decomposition in \eqref{decomposition}; the  mean current of the system is given by evaluating averages of all possible forward transitions minus all possible backward transitions with respect to the stationary probability mass function $|\boldsymbol{\rho}^{\mathrm{ss}}\rangle$ over the states of the extended state space $\mathcal{S}^{e}$. 

Using the well-known relation, $\tilde{\varphi}(\nu)= (1-\tilde{\psi}(\nu))/\nu$, between the Laplace transforms of a waiting-time density and the corresponding survival function \cite{feller}, we obtain a simple expression for the mean current  
\begin{equation}
\label{mean_j}
\langle j \rangle= \frac{r}{2} \left( \frac{\tilde{\psi}^{+}(r)}{1-\tilde{\psi}^{+}(r)} - \frac{\tilde{\psi}^{-}(r)}{1-\tilde{\psi}^{-}(r)} \right),
\end{equation}
which clearly shows a dependence of the mean current on the reorientation rate $r$. This is the chief result of this subsection and we now test it with the phase-type examples described in Sec.~\ref{typeA ratchet}. 

Applying \eqref{mean_j} for the \textcolor{black}{hypoexponential-exponential ratchet} we have
\begin{equation}
\label{hypo_mean}
\langle j\rangle = \frac{-qr}{2(p_1+p_2+r)},
\end{equation}
which is always negative, indicating that the mean current is in the backward direction. We remind the reader that we are particularly interested in the equal-mean condition for the forward and backward waiting times which for this model imposes a relation between $p_1$, $p_2$ and $q$, as given in \eqref{equal_means_hypo},  and assumed in deriving \eqref{hypo_mean}. 
For the gamma special case with \textcolor{black}{$p_1=p_2=p,\ a=2$} and the equal-mean condition, $2/p=1/q$, the mean current can easily be obtained:
\begin{equation}
\label{two_gamma_mean}
\langle j\rangle = \frac{-qr}{2(r+4q)}.
\end{equation}
On the other hand, using \eqref{mean_j} for the hyperexponential-exponential case \textcolor{black}{with the equal-mean condition \eqref{equal_mean_hyper}}, we obtain 
\begin{align}
\label{hyper_mean}
\langle j\rangle  = \frac{r}{2}\left( \frac{p_1+p_2-2q}{2r+p_1+p_2} \right),
\end{align}
which is always positive indicating that the mean current is in the forward direction.

\begin{figure}
\includegraphics[scale=0.55]{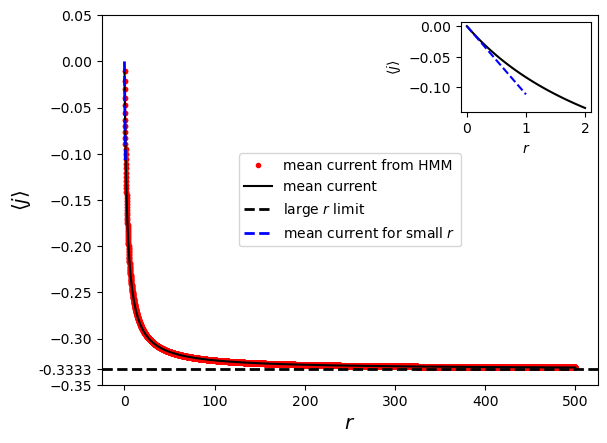}
\includegraphics[scale=0.55]{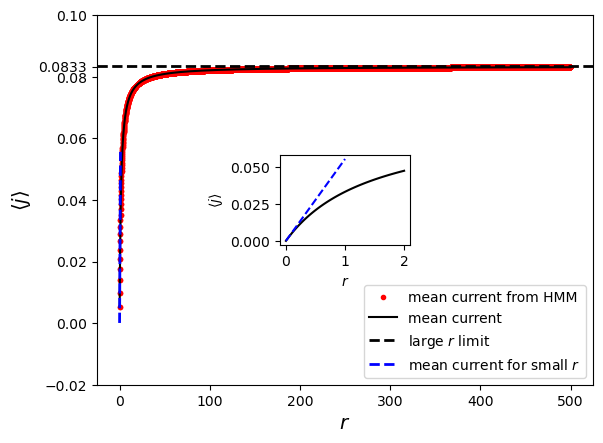}
\caption{\label{simulated_mean_currents_fig} Mean current versus reorientation rate $r$. Solid lines show the prediction of \eqref{mean_j}. Dots show the result from equivalent HMMs. Dashed black line shows the large-$r$ limit. Dashed blue line shows approximation for small $r$, with the inset displaying the behavior close to zero. Top: Mean current in the hypoexponential-exponential ratchet. A two-parameter hypoexponential distribution was used in the forward channel with parameters $p_1=1, p_2=2$. Bottom: Mean current in hyperexponential-exponential ratchet. A two-parameter hyperexponential distribution was used in the forward channel with equiprobable branches and parameters $p_1=1, p_2=2$.}
\end{figure}

The mean currents in hypoexponential-exponential and hyperexponential-exponential ratchets are plotted against the reorientation rate $r$ in Fig.~\ref{simulated_mean_currents_fig}, again with the equal-mean condition (which is assumed henceforth). The figure shows convincing agreement  with the mean current calculated from the equivalent HMMs, confirming the validity of \eqref{mean_j}. In the two examples considered, the mean-current monotonically converges to a limiting value for large $r$. A more detailed investigation of the small and large $r$ behavior of the mean ratchet current as well as its monotonicity properties is provided in the following subsections. 

\subsection{Ratchet effect}
\label{ratchet_effect}

The required condition  to see zero mean current in the system follows immediately from expression \eqref{mean_j}:
\begin{equation}
\frac{\tilde{\psi}^{+}(r)}{1-\tilde{\psi}^{+}(r)} = \frac{\tilde{\psi}^{-}(r)}{1-\tilde{\psi}^{-}(r)}.  
\end{equation}
It is then apparent that the choice
\begin{equation}
\label{condition}
\psi^{+}(\tau) \neq \psi^{-}(\tau)
\end{equation}
is both necessary and sufficient to break the symmetry in the two-channel model. Significantly, as already indicated, this implies that even if the means of both distributions are the same, a current is observed in the system when the corresponding higher moments do not match. We stress again that this is why we call the system a ratchet; indeed our condition \eqref{condition} is fully consistent with the established ``ratchet principle" that violation of time-reversal and parity symmetry leads to a directed steady-state current \cite{DENISOV201477} (see also recent discussion in \cite{ratchet_princliple_revisit}).    

Equivalently, the ratchet effect may also be inferred by investigating the generator in \eqref{effective_generator}. In general, an asymmetric $\tilde{\boldsymbol{\kappa}}(0)$ (i.e., $\tilde{\boldsymbol {\kappa}}(0) \neq [\boldsymbol{\tilde{\kappa}}(0)]^{T}$) describes an irreversible process, indicative of a system out of equilibrium with a non-zero mean current. From \eqref{effective_generator}, an asymmetric $\boldsymbol{\tilde{\kappa}}(0)$ is clearly only possible when        
\begin{equation}
\boldsymbol{\tilde{\kappa}^{+}}(0) \neq [\boldsymbol{\tilde{\kappa}^{-}}(0)]^{T},
\end{equation}
giving another method for the verification of the ratchet effect. We now proceed to further explore the general properties of this ratchet current.

\subsection{Behavior for small and large r}
\label{expansions_section}

  The analytical computation of the mean current from \eqref{mean_j} requires knowledge of probability density functions (and their Laplace transforms) for the forward and backward waiting times, which in many situations may not be available. In this subsection, we show how, without full knowledge of the waiting-time distributions, useful approximate expressions for the mean current can still be obtained when the reorientation rate $r$ is very small or very large. 

For small $r$ we can use the well-known Maclaurin  expansion for Laplace transforms of probability density functions \cite{feller,Asympt_CTRW}. 
For the forward waiting-time distribution we have
\begin{align}
\label{expansion1}
 \tilde{\psi}^{+}(r) &= 1 + r \langle \tau ^{+}\rangle + \frac{r^{2}}{2} \langle (\tau^{+})^{2} \rangle + O(r^3), 
\end{align}
 where $\langle \tau^{+}\rangle$ is the mean waiting time in the forward channel and $\langle (\tau^{+})^2\rangle$ is the corresponding second moment. With analogous notation, we have
\begin{align}
\label{expansion2}
 \tilde{\psi}^{-}(r) &= 1 + r \langle \tau ^{-}\rangle + \frac{r^{2}}{2} \langle (\tau^{-})^{2} \rangle + O(r^3)
\end{align}
for the backward waiting-time distribution. Our strategy now is to use these expansions in \eqref{mean_j} anticipating that the resulting approximation will be accurate when $r\langle \tau^{+} \rangle \ll 1$ and $r\langle \tau^{-}\rangle\ll1$, i.e., when the mean time between reorientations $1/r$ is large compared to the mean waiting times in both channels (so the particle typically makes many intra-channel moves between inter-channel moves).

Using \eqref{expansion1} and \eqref{expansion2} in \eqref{mean_j} with the equal-mean condition $\mu=\langle \tau^{+}\rangle=\langle \tau^{-} \rangle$, one finds that the leading-order behavior is linear in $r$ giving the small-$r$ approximation
\begin{equation}
\label{mean_current_approx_moments}
\langle j \rangle \approx \left ( \frac{\langle (\tau^{+})^2\rangle-\langle (\tau^{-})^2\rangle}{4\mu^{2}} \right)r.
\end{equation}
Again with the equal-mean condition this can also be written as  
\begin{equation}
\label{mean_current_approx_CV}
\langle j \rangle \approx \left(\frac{(\mathrm{CV}^{+})^2-(\mathrm{CV}^{-})^2}{4} \right) r,
\end{equation}
 where $(\mathrm{CV}^{+})^2 = \mathrm{Var}(\tau^{+})/\mu^2$ and $(\mathrm{CV}^{-})^2= \mathrm{Var}(\tau^{-})/{\mu^2}$. The coefficients of variation $\mathrm{CV}^{+}$ and $\mathrm{CV}^{-}$  provide a unitless measure of relative variance between forward and backward waiting-time distributions, which could allow convenient computations from datasets when probability density functions of the waiting times are unknown. In passing we note that one may also obtain \eqref{mean_current_approx_moments} [or equivalently \eqref{mean_current_approx_CV}] by treating the ratchet as a compound renewal process and applying renewal (reward) theorems \cite{renewal_cox}.  We will return to this renewal picture in Sec.~\ref{ratchet_large_deviations}.     

Applying the small-$r$ approximation for the mean current to the hypoexponential-exponential ratchet described in Sec.~\ref{typeA ratchet}, we obtain
\begin{equation}
\label{small_r_hypo}
\langle j \rangle \approx - \frac{p_1 p_2 r}{2(p_1 +p_2)^{2}},
\end{equation}
while, for the hyperexponential-exponential ratchet, we have
\begin{equation}
\label{small_r_hyper}
\langle j \rangle \approx  \frac{(p_1-p_2)^{2}r}{2(p_1+p_2)^2} .
\end{equation}
The small-$r$ mean currents for these ratchet systems are plotted in Fig.~\ref{simulated_mean_currents_fig} and unsurprisingly match the results computed from \eqref{mean_j} and the equivalent HMM when $r$ is close to zero. 

For large $r$, the limiting behavior of the mean current can easily be obtained from \eqref{mean_j} using $\lim_{r\to\infty}\tilde{\psi}^{\pm}(r)=0$ together with the initial value theorem $\lim_{r\to\infty} r\tilde{\psi}^{\pm}(r) =\lim_{\tau\to 0^{+}}\psi^{\pm}(\tau)$. This yields
\begin{equation}
 \langle j\rangle \approx      \frac{\psi^{+}(0)-\psi^{-}(0)}{2}
\end{equation}
which is expected to be a good approximation when $1/r$ is small compared to $\langle \tau^{+}\rangle$ and $\langle \tau^{-}\rangle$ (so the particle typically makes a maximum of one intra-channel move before an inter-channel move and Bernoulli distributions can be used in the corresponding renewal-reward process).

Using the large-$r$ result for the hypoexponential-exponential ratchet, we find that the mean current approaches $-q/2$. Intuitively, this means the directional current in this limit occurs only due to transitions in the backward Markovian channel whose rate is independent of $r$. In the $r \to \infty$ limit there is no current in the forward channel as the likelihood of forward transitions has an inverse relationship with the reorientation rate. For the hyperexponential-exponential ratchet we can also predict the limiting behavior; in the two-parameter case of \eqref{hyper_mean}, the mean current approaches $(p_1 + p_2 - 2q)/4$, showing that here transitions in both channels contribute to the current even at large $r$. The limiting behavior of these ratchet currents is confirmed by the results in Fig.~\ref{simulated_mean_currents_fig}.

\subsection{Current direction and monotonicity}
\label{mean_current_props}

In the context of ratchet systems, it is important to determine the direction of the generated current. In our setup it is apparent that the mean ratchet current depends on the waiting-time distributions in the two channels. Indeed, it is immediately obvious from \eqref{mean_j} that the mean current is in the forward direction if 
\begin{equation}
\frac{\tilde{\psi}^{+}(r)}{1-\tilde{\psi}^{+}(r)} > \frac{\tilde{\psi}^{-}(r)}{1-\tilde{\psi}^{-}(r)},
\end{equation}
and in the backward direction if
\begin{equation}
\frac{\tilde{\psi}^{+}(r)}{1-\tilde{\psi}^{+}(r)} < \frac{\tilde{\psi}^{-}(r)}{1-\tilde{\psi}^{-}(r)}.
\end{equation}

For small $r$, \eqref{mean_current_approx_CV} shows that the direction of the mean current is entirely dependent on the variances of the forward and backward process and knowledge of the full probability density functions of the waiting-time distributions is not required. In particular, if the waiting-time distribution in the backward channel is exponential, then the mean current is positive (negative) if the coefficient of variation in the forward channel is greater (smaller) than one. This crucial role of the coefficient of variation is reminiscent of other reset/renewal processes \cite{coeff_var_1, coeff_var2}.

Another interesting property to consider is the monotonicity of the generated ratchet current with respect to the reorientation rate $r$; this reveals whether there is a finite value of $r$ for which the current has a maximum magnitude. A rigorous study of the monotonicity of the general ratchet currents given by \eqref{mean_j} could perhaps be undertaken by applying the Bernstein theorem \cite{Bernstein_monotonicity} in the context of the ratio of Laplace transforms \cite{monotonicity_laplace_ratio}. However, here we simply demonstrate by example that the ratchet models discussed in this paper can be designed to give both monotone and non-monotone currents.

In the two-parameter hypoexponential-exponential and hyperexponential-exponential ratchets, we have already seen that the mean currents are monotonic with respect to the reorientation rate $r$; indeed, it is easy to check that \eqref{hypo_mean} and \eqref{hyper_mean} have no turning points where the sign of the derivatives changes with $r$. As $r$ approaches infinity, the mean currents of \eqref{hypo_mean} and \eqref{hyper_mean} saturate at finite non-zero values, as seen in Fig.~\ref{simulated_mean_currents_fig}. This implies that, in both cases, the optimal (largest magnitude) current can simply be obtained by taking the reorientation rate to infinity.

\begin{figure}
\centering
\includegraphics[scale=0.55]{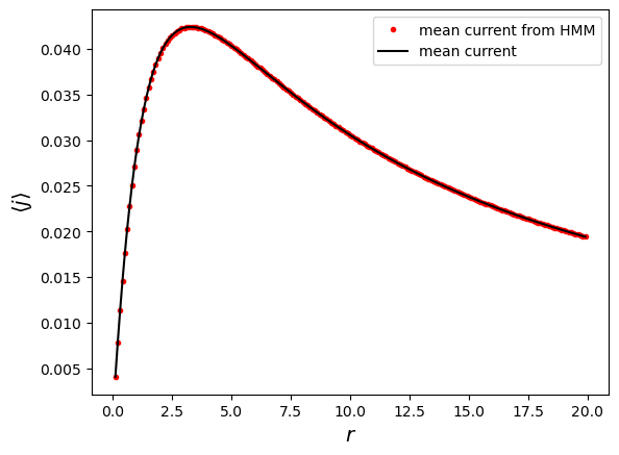}
\caption{\label{non_monotonic_plots} Non-monotonic behavior of mean current with respect to reorientation rate $r$ as seen in a ratchet with gamma-distributed waiting times in both channels. Parameters of the two gamma distributions are $(a_1, p_1)=(2,1)$ and $(a_2,p_2)=(5,2.5)$. Solid line shows the analytical mean current from \eqref{non_monotonic_gamma} while dots show results from an equivalent HMM.} 
\end{figure}

To demonstrate that the mean current can also have a non-monotonic behavior with respect to $r$ (such that some special value $r^{*}$ exists for which the mean current is optimal), we now generalize to non-exponential waiting times in both channels of the ratchet system. As a particular illustration, we consider gamma distributions for the forward and backward waiting-time densities $\psi^{+}(\tau)$ and $\psi^{-}(\tau)$, with different parameters $(a_1, p_1)$ and $(a_2, p_2)$, and the equal-mean condition $a_1/p_1 = a_2/p_2$. Taking the Laplace transform of the gamma distributions and applying \eqref{mean_j} we get
\begin{equation}
\label{non_monotonic_gamma}
\langle j \rangle =  \frac{r}{2} \left(\frac{{p_{1}}^{a_{1}}}{(r+p_{1})^{a_1}-{p_{1}}^{a_{1}}} - \frac{{p_{2}}^{a_{2}}}{(r+p_2)^{a_{2}}-{p_2}^{a_{2}}} \right).
\end{equation}
With $a_1>1$ and $a_2 > 1$, we obtain non-monotone currents as clearly shown in Fig.~\ref{non_monotonic_plots}.

\subsection{Heavy-tailed ratchets and current reversal}
\label{typeB_ratchet}

 In previous subsections we described a general method to obtain mean ratchet currents and verified it using phase-type waiting-time distributions. Now, we apply the same approach to make predictions about currents when the waiting-time distributions in both the forward and backward channels are heavy-tailed and therefore there is no equivalent hidden-Markov representation. \textcolor{black}{\textcolor{black}{CTRWs} with heavy-tailed waiting-time distributions have been studied extensively and analyzed via fractional differential equations \cite{Metzler_anomolous_diffusion, Meerschaert2018ChapterFH, gorenflo2008continuoustimerandomwalk, Hilfer_fractional_master_equations}. }

 In particular, we examine the case where  \textcolor{black}{both} waiting-time distributions have infinite mean and moments. \textcolor{black}{Here the time-averaged current in each channel, without reorientations, would tend to zero at long times and yet we shall discover that the reorientation mechanism can be exploited to generate a nonzero overall current. We thus extend our previous ratchet definition to include systems of this type, although one cannot strictly compare the (infinite) means of the heavy-tailed waiting times.} In effect, the inclusion of exponentially-distributed reorientations truncates the heavy tails, meaning that moments of the conditional waiting-time densities $W_{|\mathbf{m'}\rangle \to |\mathbf{m}\rangle}(\tau)$ are finite and hence that \eqref{mean_j} should still hold although the expansions of Sec.~\ref{expansions_section} no longer apply.

To exemplify heavy-tailed densities, we choose Mittag-Leffler distributions of the form \cite{Mittag-Leffler_1}
\begin{equation}
\label{mittag_leffler_pdf}
f_{\alpha}(\tau) = \tau^{\alpha-1} E_{\alpha,\alpha}\left[-\left(\frac{\tau}{a}\right)^{\alpha}\right] a^{-\alpha},
\end{equation}
where $a$ is a scale parameter and \textcolor{black}{$E_{\alpha,\beta}$ denotes the two-parameter Mittag-Leffler function}
\textcolor{black}
{
\begin{equation}
E_{\alpha,\beta}(z)= \sum_{n=0}^{\infty} \frac{z^{n}}{\Gamma (\alpha n+ \beta)}. 
\end{equation}
}\textcolor{black}{In the special case $\alpha=1$, the Mittag-Leffler distribution \eqref{mittag_leffler_pdf} reduces to the exponential distribution, whereas for $0<\alpha<1$, the corresponding survival function approaches a stretched   exponential for short times and $\tau^{-\alpha}$ power-law decay (infinite mean) for long times \footnote{\textcolor{black}{For more on the properties of relevant Mittag-Leffler functions, see \cite{ML_asymptotics}. CTRWs with generic power-law waiting times can be rescaled to CTRWs with Mittag-Leffler waiting-time distributions \cite{Gorenflo_mittag_leffler}. Phase behaviour and nonequilibrium stationary states induced by spatial resetting in such CTRWs and more general anomalous-diffusion processes have also been studied, see e.g., \cite{Markovian_resetting, powerlaw_resetting, anomdiff_resetting}.}}.} Significantly, the distribution has a surprisingly simple Laplace transform \cite{Mittag-Leffler_1, Mittag-Leffler2}, converging for $\nu \geq0$ to
\begin{equation}
\label{mittag_leffler_laplace}
\tilde{f}_{\alpha}(\nu)=\frac{1}{1+(a\nu)^{\alpha}}.
\end{equation}

\begin{figure}
\centering
\includegraphics[scale=0.55]{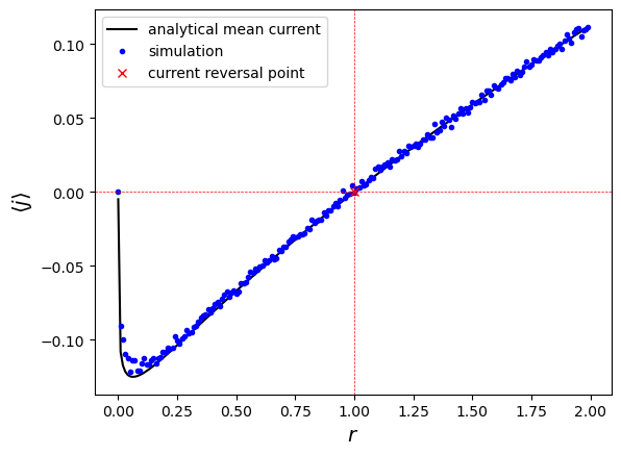}
\caption{Current reversal in a ratchet with Mittag-Leffler waiting-time distributions in the forward and backward channels with parameters $\alpha^{+}=0.5$ and $\alpha^{-}=0.75$ respectively. Solid line shows the analytical mean current from \eqref{mean_j_mittag_leffler} while dots show numerical results obtained via Gillespie Monte Carlo simulation (with $2000$ trajectories, each of duration $500$ time units). The cross shows the value of $r$ at which the mean current changes direction.}
\label{current_reversal_fig}
\end{figure}

In the forward channel of the system described in Sec.~\ref{Model}, we now choose the Mittag-Leffler waiting-time distribution
\begin{align}
\psi^{+}(\tau) = \tau^{\alpha^{+}-1} E_{\alpha^{+},\alpha^{+}}\left[-\left(\frac{\tau}{a}\right)^{\alpha^{+}}\right]a^{-\alpha^{+}},
\end{align}
with $\alpha^{+}<1$, while in the backward channel we take
\begin{align}
\psi^{-}(\tau) = \tau^{\alpha^{-}-1} E_{\alpha^{-},\alpha^{-}}\left[-\left(\frac{\tau}{a}\right)^{\alpha^{-}}\right]a^{-\alpha^{-}},
\end{align}
with  $\alpha^{-} < 1$. We set $\alpha^{+} \neq \alpha^{-}$, such that the tails of the waiting-time distributions decay at different rates in the forward and backward channels, and fix $a=1$ for convenience. Using the Laplace transform \eqref{mittag_leffler_laplace} in \eqref{mean_j}, we have the mean current
\begin{equation}
\label{mean_j_mittag_leffler}
\langle j \rangle = \frac{r^{1-\alpha^{+}}-r^{1-\alpha^{-}}}{2},
\end{equation}
which interestingly shows a current reversal at $r=1$. As shown in Fig.~\ref{current_reversal_fig}, with $\alpha^{+}<\alpha^{-}$ the mean current is negative for $r<1$ and positive for $r>1$. Here, there is no equivalent HMM but the results are confirmed via Monte Carlo simulation.

\section{Fluctuations}
\label{ratchet_large_deviations}
\subsection{Large-deviation framework}
The large-deviation approach to nonequilibrium systems \cite{rosemary_chapter, touchette1} defines key quantities analogous to the free energy and the entropy in equilibrium statistical physics: the \textit{scaled cumulant generating function} (SCGF) and the \textit{rate function}. These characterize fluctuations of time-averaged observables and are therefore relevant for understanding rare events (atypical behavior) of time-averaged quantities, such as the time-averaged ratchet current $j$ of interest in this work.

The large-deviation principle, generically found  in Markovian systems is governed by a rate function $I(\j)$, and can be loosely written as 
\begin{equation}
\label{LDP}
\mathbb{P} \left( j(t) =\j \right) \sim e^{-tI(\j)},
\end{equation}
where the symbol $\sim$ signifies logarithmic equality in the long-time limit. The rate function here describes the exponentially-decaying probability of rare events; the $t$ appearing in the exponent is sometimes known as the \textit{speed} of the large deviations. For currents in non-Markovian systems, there may be no large-deviation principle or one with a modified speed (see for example \cite{ Harris_erw, jack_harris}). However, for the ratchet models considered in this paper there is, in fact, a large-deviation principle of the standard form \eqref{LDP} and we seek to investigate the properties of fluctuations encoded in $I(\j)$.

The rate function is related to the SCGF $\lambda(s)$ via the G\"artner-Ellis theorem: 
\begin{equation}
 \label{GE}
 I(\j) = \sup_{s} \{ s\j - \lambda(s) \}.          
 \end{equation}
 The SCGF in turn is defined as 
\begin{equation}
 \label{SCGF1}
 \lambda(s) = \lim_{t\to \infty} \frac{1}{t}\ln \mathbb{E}[e^{sJ(t)}],  
 \end{equation}
where $\mathbb{E}[e^{sJ(t)}]$ is the generating function of the time-integrated current and plays the role of a thermodynamic partition function. Note that the expectation here is over all trajectories of the process and, for long times and ergodic dynamics, is equivalent to a steady-state average.

The SCGF is often easier to calculate than the rate function and, for Markov systems, can be calculated analytically via a spectral method; $\lambda(s)$ is the principal eigenvalue of the tilted generator matrix. For the phase-type semi-Markov processes of Sec. \ref{typeA ratchet} the SCGF can be obtained from the tilted generator matrix of the equivalent HMM but this method is generally not applicable for other non-Markov systems. However, it turns out that for the specific ratchet models considered in this paper (with arbitrary waiting-time distributions) the SCGF can be semi-analytically derived via a framework using renewal theory, as we now explain. Henceforth, for definiteness, we consider a setup where the particle has just entered the forward channel at $t=0$.

\subsection{SCGF from renewal theory}
\label{SCGF_from_renewal_theory}

 The crucial observation here is that our ratchet model may be viewed as a renewal process with cycles composed of a forward run of duration $t^{+}$ and a backward run of duration $t^{-}$, such that the length of the whole cycle is $t^{+}+t^{-}$. By construction, $t^{+}$ and $t^{-}$ are random variables from an exponential distribution with rate $r$. The runs are themselves  renewal processes  with inter-event times from the waiting-time distributions, $\psi^{+}(\tau)$ and $\psi^{-}(\tau)$.


To calculate the SCGF, we begin by computing the moment generating functions $U(s,t^{+})$ and $V(s,t^{-})$ for currents in the forward and backward runs. For the forward run, we have 
\begin{equation}
U(s,t^{+})= \mathbb{E}^{+}[{e^{sJ^{+}(t^{+})}}],
\end{equation}
where this is the expectation of the weighted exponential number of jumps in the forward channel in the interval between entering the channel and leaving it a time $t^{+}$ later.
In principle, to calculate this expectation one needs to average over trajectory segments with $n=0,1,2 \dots $ jumps, while imposing the condition that the sum of the $n$ waiting times and the final survival time is equal to $t^{+}$. It is easier to relax this constraint and perform the computation in the Laplace domain, where we have the transformed generating function
\begin{align}
\nonumber \tilde{U}(s,\nu) =\tilde{\varphi}^{+}(\nu) &+ e^{s}\tilde{\psi}^{+}
(\nu)\tilde{\varphi}^{+}(\nu)\\& +e^{2s}\tilde{\psi}^{+}(\nu)\tilde{\psi}^{+}(\nu) \tilde{\varphi}^{+}(\nu)+\cdots.
\end{align}
Here the first term corresponds to a forward run with no jumps, the second term corresponds to a forward run with one jump, the third term corresponds to a forward run with two jumps, and so on. Recognizing the sum as a geometric series, we finally get
\begin{align}
\nonumber \tilde{U}(s,\nu) &=\frac{\tilde{\varphi}^{+}(\nu)}{1-e^{s}\tilde{\psi}^{+}(\nu)} \\ &= \frac{1-\tilde{\psi}^{+}(\nu)}{\nu(1-e^{s}\tilde{\psi}^{+}(\nu))}.
\end{align}
Similarly, for the backward channel, where the current is negative by construction, we have
\begin{align}
 \tilde{V}(s,\nu) = \frac{1-\tilde{\psi}^{-}(\nu)}{\nu(1-e^{-s}\tilde{\psi}^{-}(\nu))} .
\end{align}

To construct the moment generating function $G_{r}(s,t)$ of the overall ratchet system , we need to weight the generating functions $U(s,t^{+})$ and $V(s,t^{-})$ by the probabilities of seeing runs of length $t^{+}$ and $t^{-}$. To this end, we define
\begin{equation}
G^{+}_{r}(s,t^{+})=  re^{-rt^{+}} U(s,t^{+}) 
\end{equation}
and find its Laplace transform
\begin{align}
\label{mgf_int1}
\tilde{G}^{+}_r(s,\nu) &=\int_{0}^{\infty}e^{-\nu t^{+}} r e^{-rt^{+}} U(s,t^{+}) \mathrm{d}t^{+}\\  &= r \tilde{U}(s,\nu+r) \\ 
\label{mgf_plus}
& = \frac{r(1-\tilde{\psi}^{+}(\nu+r))}{(\nu+r)(1-e^{s}\tilde{\psi}^{+}(\nu+r))}.  
\end{align}
Similarly, of course, we have 
\begin{align}
 G^{-}_{r}(s,t^{-})= re^{-rt^{-}} V(s,t^{-}) 
\end{align}
and
\begin{align}
\label{mgf_int_2}
 \tilde{G}_{r}^{-}(s,\nu) &=\int_{0}^{\infty}e^{-\nu t^{-}} r e^{-rt^{-}} V(s,t^{-}) \mathrm{d}t^{-} \\ 
 \label{mgf_minus}
 & =\frac{r(1-\tilde{\psi}^{-}(\nu+r))}{(\nu+r)(1-e^{-s}\tilde{\psi}^{-}(\nu+r))}.
\end{align}

The overall ratchet is composed of alternating segments of forward and backward runs. Since we are interested in the long-time behavior, and the $t^{\pm}$ have finite moments, we consider only a complete number of renewal cycles meaning that in the Laplace domain we have a simple geometric sum 
\begin{align}
\label{geometric_sum_mgf}
\nonumber \tilde{G}_{r}(s,\nu) &= \tilde{G}^{+}_{r}(s,\nu)\tilde{G}_{r}^{-}(s,\nu) + (\tilde{G}^{+}_{r}(s,\nu) \tilde{G}_{r}^{-}(s,\nu))^{2}+ \cdots \\ 
& =\frac{\tilde{G}_{r}^{+}(s,\nu)  \tilde{G}_{r}^{-}(s,\nu)}{1-\tilde{G}_{r}^{+}(s,\nu) \tilde{G}_{r}^{-}(s,\nu)}.
\end{align}
Note that allowing for incomplete cycles would produce temporal boundary terms, modifying only the numerator of this equation. 

As pointed out, for example in \cite{phase_transitions}, the form  of \eqref{geometric_sum_mgf} is analogous to the spatial partition function used by Poland and Scheraga to study phase transitions in a DNA denaturation model \cite{poland-scheraga, poland_scheraga_2}.  In a similar spirit to the original Poland-Scheraga analysis, the SCGF here is given by the largest real value of $\nu$ where $\tilde{G}_{r}(s,\nu)$ becomes infinite (i.e., the convergence boundary point). From \eqref{geometric_sum_mgf}, we see that $\tilde{G}_{r}(s,\nu)$ can become infinite when (i) $\tilde{G}_{r}^{+}(s,\nu) \tilde{G}^{-}_{r}(s,\nu)=1$, (ii)  $\tilde{G}_{r}^{+}(s,\nu)$ becomes infinite, or (iii) $\tilde{G}_{r}^{-}(s,\nu)$ becomes infinite. We denote the largest real value of $\nu$,  corresponding to each of these three conditions by $\nu_{*}$, $\nu^{+}_{*}$ and $\nu^{-}_{*}$. 

\begin{figure}
\centering
\includegraphics[scale=0.6]{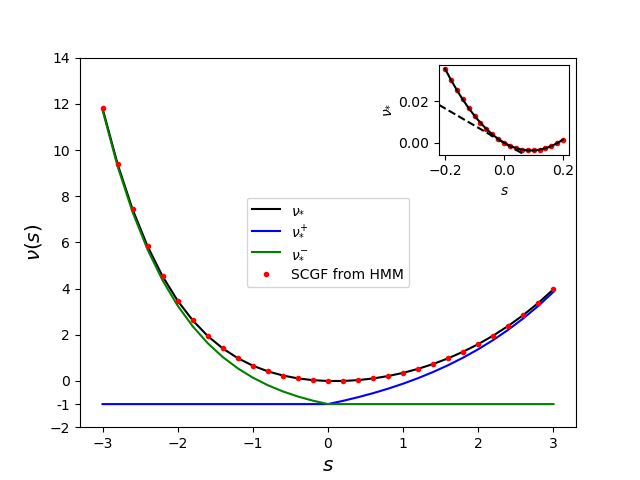}
\includegraphics[scale=0.6]{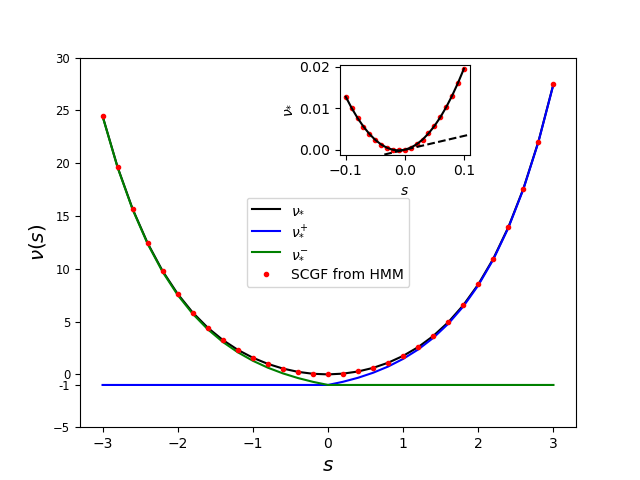}
\caption{Analysis of SCGF in phase-type ratchets.  Lines show values of $\nu$ where $\tilde{G}_{r}(s,\nu)$ diverges: $\nu_{*}$ (black), $\nu^{+}_{*}$ (blue) and $\nu^{-}_{*}$ (green). Dots show SCGF computed semi-analytically with equivalent HMMs. Top: two-parameter hypoexponential-exponential ratchet with $p_1=1, p_2=2$ and reorientation rate $r=1$. Bottom: two-parameter hyperexponential-exponential ratchet with $p_1=1$, $p_2=2$ and reorientation rate $r=1$. \textcolor{black}{Insets show the plots of $\nu_{*}$ (which in this case gives the SCGF) zoomed-in about $s=0$ with slopes of the dotted lines corresponding to the mean currents.}}
\label{fig_phase_type_ratchet_SCGF_analysis}
\end{figure}

To see how this works in practice, we plot in Fig.~\ref{fig_phase_type_ratchet_SCGF_analysis} the numerically-determined values $\nu_{*}$, $\nu^{+}_{*}$ and $\nu^{-}_{*}$ for the phase-type ratchets of Sec.~\ref{typeA ratchet}. In both cases $\nu_{*}$ is the largest
 (approaching $\nu^{^+}_{*}$ and $\nu^{-}_{*}$ asymptotically), so this gives directly the SCGF. Note that the slope of the SCGF at $s=0$ gives the mean current: negative for the hypoexponential-exponential ratchet and positive for the hyperexponential-exponential ratchet. The renewal theory results are unsurprisingly in full agreement with the SCGF obtained from the tilted generator matrices of equivalent HMMs (shown in the appendix \ref{tilted_matrices}). They have also been confirmed via a recent trajectory-based reinforcement learning algorithm \cite{Pamulaparthy_2025}. 

\begin{figure}
\includegraphics[scale=0.6]{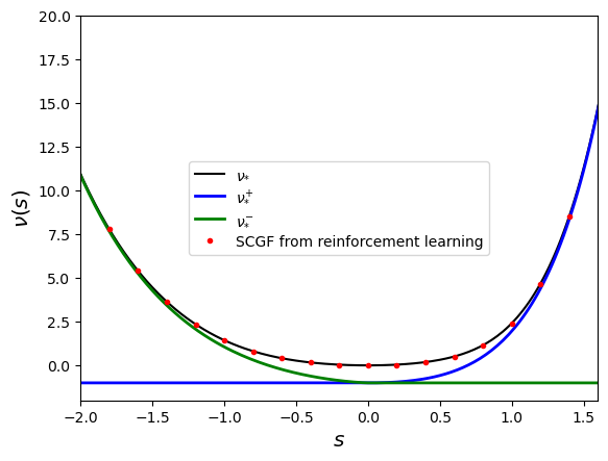}
\caption{Same as Fig.~\ref{fig_phase_type_ratchet_SCGF_analysis} but for heavy-tailed ratchet with Mittag-Leffler waiting-time distributions in the forward and backward channels: $\alpha^{+}=0.5$, $\alpha^{-}=0.75$. Dots here show results using the reinforcement learning algorithm of \cite{Pamulaparthy_2025}.} 
\label{fig_mittag_leffler_SCGF_analysis}
\end{figure}

In Fig.~\ref{fig_mittag_leffler_SCGF_analysis} we show the corresponding analysis for a ratchet with heavy-tailed waiting times for intra-channel moves but still with exponential reorientations (as described in Sec.~\ref{typeB_ratchet}). An equivalent HMM is not available for verification in this case so we show instead results from the reinforcement learning algorithm \cite{Pamulaparthy_2025}. The picture is similar to Fig.~\ref{fig_phase_type_ratchet_SCGF_analysis}, but the line for $\nu_{*}$ becomes extremely close to those of $\nu^{+}_{*}$ and $\nu_{*}^{-}$  (for large positive $s$ and large negative $s$ respectively) and it is not immediately clear by inspection if they actually touch. The lines touching would indicate that there exist dynamical phase transitions; we investigate this possibility in the next subsection. 

\subsection{Dynamical phase transitions}
\label{phase_transition_section}
As noted above, the Laplace-transformed moment generating function of the full ratchet system given in \eqref{geometric_sum_mgf} can only diverge when $\tilde{G}_{r}^{+}(s,\nu)$ diverges, $\tilde{G}^{-}_{r}(s,\nu)$ diverges or $\tilde{G}_{r}^{+}(s,\nu)\tilde{G}_{r}^{-}(s,\nu)=1$. If, for some $s$, the largest value of $\nu$ for which $\tilde{G}_{r}^{+}(s,\nu)\tilde{G}_{r}^{-}(s,\nu)=1$ coincides with the convergence boundary point of $\tilde{G}_{r}^{+}(s,\nu)$ or $\tilde{G}^{-}_{r}(s,\nu)$ (i.e., the lines for $\nu_{*}$ and $\nu_{*}^{\pm}$ cross or touch when plotted against $s$) then this gives a non-analytic point in $\lambda(s)$ corresponding to a dynamical phase transition. Physically, this indicates that fluctuations in different regimes are controlled by different mechanisms. Such a scenario cannot happen for phase-type ratchets since these systems are Markovian in the extended HMM state space, which means that the spectrum of their tilted generator matrix will always be gapped (by a Perron-Frobenius--type argument). We now seek to determine whether there are conditions for which dynamical phase transitions do happen in other ratchet systems. 



We start by looking carefully at the possibility of intersection of $\nu_{*}$ and $\nu^{+}_{*}$. Now, the Laplace-transformed moment generating function of current in the forward channel $\tilde{G}_{r}^{+}(s,\nu)$ diverges when either the numerator of \eqref{mgf_plus} becomes infinite or the denominator becomes zero. Since, by normalization, the probability density function $\tilde{\psi}^{+}(\nu+r)$ must converge at $\nu+r=0$, we note that any divergence in the numerator can only occur at $\nu<-r$. On the other hand, the denominator is zero when $\nu=-r$ or when $1-e^{s}\tilde{\psi}^{+}(\nu+r)=0$, with these two poles crossing over at $s=0$.  We thus deduce that $\nu^{+}_{*}$, the convergence boundary point of $\tilde{G}_{r}^{+}(s,\nu)$, always corresponds to the largest pole in the denominator (which must be at least $-r$) rather than divergence of the numerator. Significantly, this means that at the convergence boundary, $\tilde{G}_{r}^{+}(s,\nu^{+}_{*})$ is infinite and cannot simultaneously satisfy the condition \textcolor{black}{ $\tilde{G}_{r}^{+}(s,\nu)\tilde{G}_{r}^{-}(s,\nu)=1$}.

Note that the argument of the preceding paragraph applies for any normalizable $\psi^{+}(\tau)$, including with heavy tails, and shows that $\nu_{*}$ and $\nu^{+}_{*}$ can never intersect. A similar argument, of course, precludes the intersection of $\nu_{*}$ and $\nu^{-}_{*}$. In other words, for any intra-channel waiting-time distributions with exponential reorientation times, it is impossible to see dynamical phase transitions; even in Fig.~\ref{fig_mittag_leffler_SCGF_analysis} the lines do not actually touch.

In a continued search for more interesting phenomenology, we conclude the paper with a brief treatment of more exotic ratchets where the reorientation mechanism is non-Markovian. In fact, it is easy to show, by the same argument as above, that there are no dynamical phase transitions for any light-tailed distribution of reorientation times. However, it turns out that for heavy-tailed reorientation distributions, dynamical phase transitions \emph{are} possible as we now illustrate explicitly with a simple example for which analytical calculations can be carried out.

To be concrete, we choose a Mittag-Leffler distribution with  $0<\alpha<1$ (and $a=1$) for the reorientation times while for transitions in the forward and backward channels we now take exponential waiting times with rate $q$. In other words, the intra-channel moves are now Markovian but the inter-channel moves (reorientations) are non-Markovian. In this case, \eqref{mgf_int1} and \eqref{mgf_int_2}, are replaced by
\begin{equation}
\label{mgf_phase_trans_plus}
\tilde{G}_{\alpha}^{+}(s,\nu)= \int_{0}^{\infty} e^{-\nu t^{+}}e^{q(e^{s}-1)t^{+}} f_{\alpha}(t^{+})\mathrm{d}t^{+} 
\end{equation}
and
\begin{equation}
\label{mgf_phase_trans_minus}
\tilde{G}_{\alpha}^{-}(s,\nu)= \int_{0}^{\infty} e^{-\nu t^{-}} e^{q(e^{-s}-1)t^{-}} f_{\alpha}(t^{-})\mathrm{d}t^{-},
\end{equation}
where $f_{\alpha}(t)$ is the Mittag-Leffler probability density function  \eqref{mittag_leffler_pdf} while $e^{q(e^{s}-1)t^{+}}$ and $e^{q(e^{-s}-1)t^{-}}$ are moment generating functions of the Poisson currents in the forward and backward channels respectively. The exponential nature of the moment generating functions enables the use of the shifting property of Laplace transforms. From \eqref{mittag_leffler_laplace} we have that $\tilde{G}_{\alpha}^{+}(s,\nu)$ converges for $\nu\geq \nu^{+}_{*}=q(e^{s}-1)$ to
\begin{equation}
\tilde{G}^{+}_{\alpha}(s,\nu)= \frac{1}{1+(\nu-q(e^{s}-1))^{\alpha}}.
\end{equation}
Similarly, for the backward channel, $\tilde{G}_{\alpha}^{-}(s,\nu)$ converges for $\nu\geq \nu^{-}_{*}=q(e^{-s}-1)$ to
\begin{equation}
\tilde{G}^{-}_{\alpha}(s,\nu)= \frac{1}{1+(\nu-q(e^{-s}-1))^{\alpha}}.
\end{equation}

Since we now have heavy-tailed waiting times we should also consider incomplete renewal cycles leading to additional terms in the numerator of \eqref{geometric_sum_mgf}. These terms involve analogs of \eqref{mgf_phase_trans_plus} and \eqref{mgf_phase_trans_minus} with the probability density replaced by the corresponding survival function. However, it is easy to check (again due to the Poisson generating functions) that these terms do not introduce additional divergences into the overall Laplace-transformed generating function $\tilde{G}_{\alpha}(s,\nu)$.

\begin{figure}
\centering
\includegraphics[scale=0.55]{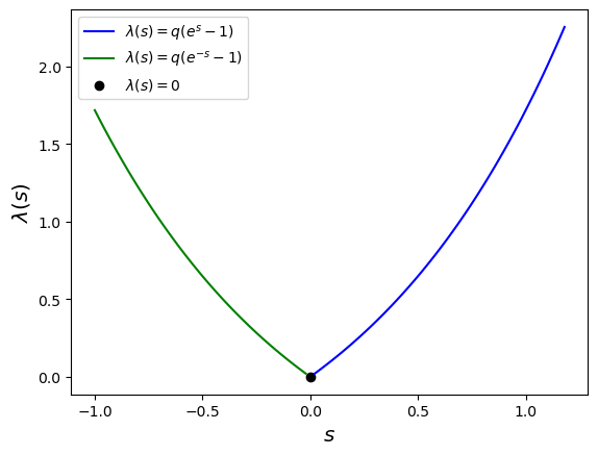}
\caption{SCGF of the ratchet model with Mittag-Leffler reorientation mechanism ($0<\alpha<1$) and exponential waiting-time distributions ($q=1$) in the forward and backward channels. Green and blue lines show SCGF for $s<0$ and $s>0$ respectively; black dot indicates the first-order phase transition at $s=0$. }
\label{phase_transition_ratchet}
\end{figure}

Significantly, in contrast to the case with light-tailed reorientation distributions, $\tilde{G}_{\alpha}^{+}(s,\nu)$ and $\tilde{G}_{\alpha}^{-}(s,\nu)$ are finite at their convergence boundaries and hence it may be possible to simultaneously satisfy the condition $\tilde{G}_{\alpha}^{+}(s,\nu)\tilde{G}_{\alpha}^{-}(s,\nu)=1$. Indeed, we see by inspection here that $\tilde{G}_{\alpha}^{+}(s,\nu^{+}_{*})=1$ and $\tilde{G}_{\alpha}^{-}(s, \nu^{-}_{*})=1$, and that $\nu^{+}_{*}$, $\nu_{*}^{-}$ and $\nu_{*}$ all coincide at $s=0$. 
This leads to the piecewise SCGF  
\begin{align}
\lambda(s)= \begin{cases} q(e^{-s}-1) , & \text{for $s<0$}\\
                0,  &\text{for $s=0$} \\
             q(e^{s}-1)  & \text{for $s>0$},
            \end{cases}
\end{align}
as plotted in Fig.~\ref{phase_transition_ratchet}. Notice that the tail parameter $\alpha$ does not appear here (assuming $0<\alpha<1$).

\begin{figure}
\centering
\includegraphics[scale=0.55]{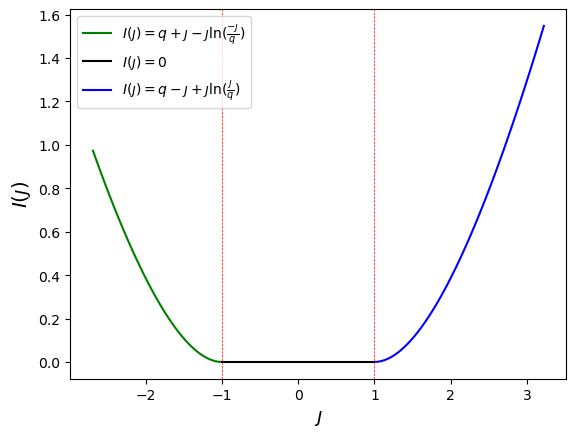}
\caption{Rate function of the ratchet model with Mittag-Leffler reorientation mechanism ($0<\alpha<1$) and exponential waiting-time distributions ($q=1$) in the forward and backward channels. Colors of the lines indicate the mapping from the SCGF of Fig.~\ref{phase_transition_ratchet}.}
\label{ratchet_rate_function}
\end{figure}

There exists now a non-analytic point in the SCGF $\lambda(s)$ at $s=0$, indicating a dynamical phase transition in the fluctuations. In particular, since the first derivative is not continuous, this is a first-order phase transition. Via the Legendre-Fenchel transform \eqref{GE}, we obtain the rate function  \begin{align}
\label{rate_function_eq}
I(\j)= \begin{cases} q+\j-\j\ln(-\frac{\j}{q}) , &\text{for $\j<-q$}\\
                0,  & \text{for $-q\leq \j \leq q$}\\
                q-\j+\j\ln(\frac{\j}{q}) &\text{for $\j>q$},
            \end{cases}
\end{align}
where the single point at $s=0$ in the SCGF (black dot in Fig.~\ref{phase_transition_ratchet}) maps to a horizontal line in the rate function (black line in Fig.~\ref{ratchet_rate_function}).

Mathematically, since the SCGF is non-differentiable the Legendre-Fenchel transform only gives the convex hull of the rate function. However, we can argue on physical grounds that \eqref{rate_function_eq} is indeed the true rate function. Intuitively, the heavy-tails in the reorientation distribution allow the process to remain for an arbitrarily long time in one channel without any probabilistic cost (on the large-deviation scale); \textcolor{black}{this is a nonergodic system without a unique stationary state}. This implies that \emph{any} time-averaged current between $-q$ and $+q$ can be achieved equiprobably by the system spending a macroscopic fraction of its history in the forward channel (with mean current $+q$) and the remainder in the backward channel (with mean current $-q$). The phase separation in time is analogous to the Maxwell construction, characteristic of first-order phase transitions in equilibrium systems. The fact that the system has a range of equiprobable currents on the exponential scale means that an individual trajectory will typically display a non-zero current and in this sense the system can be thought of as a ratchet even though the rate function of Fig.~\ref{ratchet_rate_function} is symmetrical.

We highlight that on either side of the phase transition the fluctuations are controlled by different mechanisms: time-averaged currents greater than $q$ are optimally realized by remaining in the forward channel while time-averaged currents less than $-q$ are optimally realized by remaining in the backward channel. This competition between mechanisms is also anticipated to lead to similar dynamical phase transitions in other ratchet models with heavy-tailed reorientation times.

\section{Discussion}
\label{discussion}
In this work we introduced a novel type of ratchet mechanism where non-zero currents are formed in a two-channel system (reminiscent of run-and-tumble dynamics) due to an asymmetry in waiting-time distributions. This ratchet is in the spirit of classical resetting models  \cite{resetting_sabhapandit, Evans_resetting}, but instead of transitioning to an earlier configuration or current,  the reset mechanism transfers the dynamics to the opposite channel. 

For the bulk of the paper, we concentrated on the case where the reorientation between channels is governed by an exponential distribution. Here, we showed that even if the waiting-time distributions in the forward and backward channels have the same mean it is possible to engineer positive or negative ratchet currents \footnote{\textcolor{black}{Note that if the hopping in both channels were in the same direction, then switching could give an overall mean current larger than the natural (nonswitching) current in either channel; this is effectively a realisation of Parrondo's paradox \cite{Parrondo}.}}; we derived an explicit formula for the mean ratchet current in terms of Laplace-transformed distributions. Furthermore, we were able to go beyond the typical behavior by developing a framework for computing the ratchet-current SCGF via renewal theory. We argued that for exponential reorientations (or indeed any light-tailed distribution for the inter-channel moves) there are no dynamical phase transitions; fluctuations away from the mean always involve contributions from both forward and backward channels.



Finally, we applied the renewal framework to study a particular example of a ratchet with a heavy-tailed distribution of reorientation times. This case manifests a first-order dynamical phase transition in which large forward currents are optimally realised by the system remaining in the forward channel while large backward currents are optimally realised by remaining in the backward channel. Between these two phases, there is a coexistence regime for intermediate fluctuations, suggesting a whole range of typical currents that are equiprobable on the exponential scale. It would be interesting to investigate further structure within this regime (e.g., under a different scaling) as well as to look for examples of ratchet models which have higher-order dynamical phase transitions and therefore no phase coexistence. 

Significantly, we note that external potentials are not needed to generate currents in our model so it could be relevant for the understanding of active particle systems such as molecular motors \cite{ratchet_molecular_motors}. In practice, it may be difficult to control waiting-time distributions to produce a desired current but we could perhaps shed light on the inverse problem: if for some biological-transport situation the current is observed to change with the tumble frequency, non-Markovian properties of the underlying dynamics can be inferred. Extensions to the quantum realm and related applications also offer an interesting direction to pursue \cite{quantum_ratchets, reset_closed_quantum, reset_for_open_quantum}.

\begin{acknowledgments}
VDP is grateful for feedback and suggestions from Federico Carollo and Fabio Caccioli; he also thanks Arvind Ayyer and the Indian Institute of Science for hospitality during the final stages of manuscript preparation. RJH \textcolor{black}{is thankful to Ion Santra, Martin Weigel, and Philipp Maass for insightful comments, and further} acknowledges support from the London Mathematical Laboratory.
\end{acknowledgments}
\textcolor{black}{
\section*{data availability}
The data and code that support the findings of this work are openly available \footnote{{h}ttps://github.com/Venkata-Dhruva-Pamulaparthy/ratchets.git}.
}

\appendix*



\begin{widetext}
\section{Tilted generator matrices for phase-type ratchets}
\label{tilted_matrices}

We outline here the spectral approach for computing the SCGF in the two-parameter hypoexponential-exponential and hyperexponential-exponential ratchets described in Sec.~\ref{typeA ratchet} of the main text. Since the current fluctuations are unaffected by the system size, for convenience we choose a ring of three-sites so that the generator of the equivalent HMM can be represented by a $9 \times 9$ matrix. To obtain the \textit{tilted} generator, every element corresponding to a forward move is weighted by an exponential factor $e^{s}$ while every element corresponding to a backward move is weighted by $e^{-s}$. For the hypoexponential-exponential ratchet on a three-site ring, we  obtain the tilted generator 
\begin{align}
\label{tilted_hypo}
{\mathrm{\mathbf{G}}}(s) =  \begin{bmatrix} -p_{1}-r & 0 & 0 & 0 & 0 & p_2 e^{s}  & r & 0 & 0 \\  p_1 & -p_{2}-r & 0 & 0 & 0 & 0 &0  &0 &0\\ 0 & p_2e^{s} & -p_{1}-r & 0 & 0 & 0 &0 & 0 &0  \\  0 & 0 & p_1& -p_{2}-r & 0 & 0 &0 & 0 &0\\ 0 & 0 & 0 & p_{2}e^{s}  & -p_{1}-r & 0&0 &0  &0 \\ 0 & 0 & 0 & 0 & p_1 & -p_2-r&0 &0  &0\\ r & r& 0 &0 &0 &0 & - q-r & qe^{-s} & 0 \\ 0 & 0& r & r& 0 &0 &0 & -q-r & qe^{-s}\\  0 &0 &0 &0 &r & r &qe^{-s} &0 & -q
\end{bmatrix}
\end{align}
and for the hyperexponential-exponential ratchet we get
\begin{align}
\label{tilted_hyper}
{\mathrm{\mathbf{G}}}&(s) =
\begin{bmatrix} -p_{1}-r & 0 & 0 & 0 & p_1 e^{s}/2 & p_2e^{s}/2 & r/2 & 0 & 0 \\  0 & -p_{2}-r& 0& 0 & p_1 e^{s}/2 & p_2 e^{s}/2 & r/2 & 0 & 0 \\ p_1e^{s}/2 & p_2e^{s}/2 & -p_{1}-r & 0 & 0 & 0& 0& r/2 & 0 \\  p_{1} e^{s}/2 & p_2e^{s}/2 & 0 & -p_{2}-r & 0 & 0 & 0 & r/2 & 0 \\ 0 & 0 & p_1e^s/2 & p_2e^{s}/2& -p_{1}-r & 0 & 0 & 0 & r/2 \\ 0 & 0  & p_{1}e^{s}/2& p_2e^{s}/2& 0& -p_{2}-r & 0 & 0 & r/2 \\ r & r & 0 & 0 & 0 & 0 & -q-r & qe^{-s} & 0 \\ 0 & 0 & r & r & 0 & 0 & 0 & -q-r & q e^{-s} \\ 0 & 0 & 0 & 0 & r & r & qe^{-s} & 0& -q-r \end{bmatrix}.
\end{align}
The SCGF is then given by the principal eigenvalue of the corresponding tilted generator \cite{touchette1,rosemary_chapter}.

\end{widetext}

\nocite{*}

%

\end{document}